\documentclass[journal]{IEEEtran}
\ifCLASSINFOpdf
\usepackage[pdftex]{graphicx}
\graphicspath{{./image/}}
\DeclareGraphicsExtensions{.pdf,.jpeg,.png}
\else
\fi

\usepackage{latexsym}
\usepackage{amssymb}
\usepackage[cmex10]{amsmath}
\usepackage{array}
\ifCLASSOPTIONcompsoc
  \usepackage[caption=false,font=normalsize,labelfont=sf,textfont=sf]{subfig}
\else
  \usepackage[caption=false,font=footnotesize]{subfig}
\fi
\begin{document}

\title{Unsupervised Learnable Sinogram Inpainting Network (SIN) for Limited Angle CT reconstruction}

\author{Ji~Zhao, Zhiqiang~Chen* , Li~Zhang, Xin~Jin
\thanks{This work was supported by a grant from the National Natural Science Foundation of China (No. 11235007), "Research on X-ray Imaging Theory and Key Technologies".

  Ji Zhao, Zhiqiang Chen and Li Zhang are with Key Laboratory of Particle and Radiation Imaging (Tsinghua University), Ministry of Education, Beijing 100084, China and the Department of Engineering Physics, Tsinghua University, Beijing 100084, China, and Xin Jin iswith Smart Inspection Division, Nuctech Company Limited.(Corresponding author: Zhiqiang Chen, czq@tsinghua.edu.cn)}
}

\maketitle

\begin{abstract}
In this paper, we propose a sinogram inpainting network (SIN) to solve
limited-angle CT reconstruction problem, which is a very challenging ill-posed
issue and of great interest for several clinical applications. A common approach
to the problem is an iterative reconstruction algorithm with regularization
term, which can suppress artifacts and improve image quality, but requires high
computational cost.

The starting point of this paper is the proof of inpainting function for
limited-angle sinogram is continuous, which can be approached by neural networks
in a data-driven method, granted by the universal approximation theorem. Based
on this, we propose SIN as the fitting function – a convolutional neural network
trained to generate missing sinogram data conditioned on scanned data. Besides
CNN module, we design two differentiable and rapid modules, Radon and Inverse
Radon Transformer network, to encapsulate the physical model in the training
procedure. They enable new joint loss functions to optimize both sinogram and
reconstructed image in sync, which improved the image quality significantly. To
tackle the labeled data bottleneck in clinical research, we form a
sinogram-image-sinogram closed loop, and the difference between sinograms can be
used as training loss. In this way, the proposed network can be self-trained,
with only limited-angle data for unsupervised domain adaptation.

We demonstrate the performance of our proposed network on parallel beam X-ray CT
in lung CT datasets from Data Science Bowl 2017 and the ability of unsupervised
transfer learning in Zubal's phantom. The proposed method performs better than
state-of-art method SART-TV in PSNR and SSIM metrics, with noticeable visual
improvements in reconstructions.
\end{abstract}


\IEEEpeerreviewmaketitle

\section{Introduction}
\label{sec:orgb798636}

\begin{figure}[tb]
\centering
\subfloat[]{\includegraphics[width=1.1in]{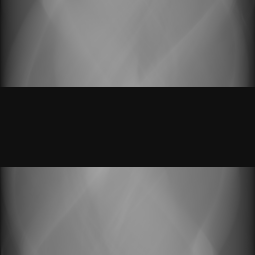}}
\hfill
\subfloat[]{\includegraphics[width=1.1in]{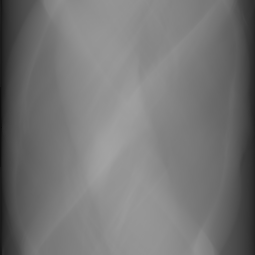}}
\hfill
\subfloat[]{\includegraphics[width=1.1in]{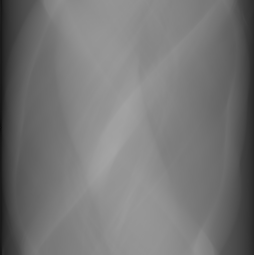}}
\\
\subfloat[]{\includegraphics[width=1.1in]{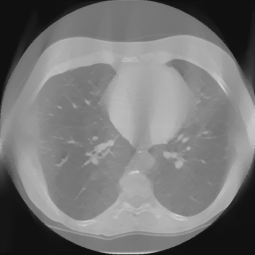}}
\hfill
\subfloat[]{\includegraphics[width=1.1in]{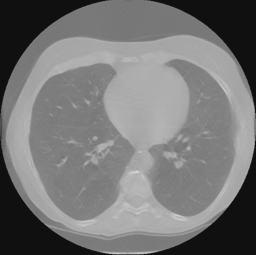}}
\hfill
\subfloat[]{\includegraphics[width=1.1in]{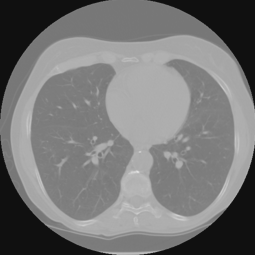}}
\\
\caption{Qualitative illustration of the task: (a) given a limited-angle sinogram; (b) automatic inpainting result using our sinogram inpainting network (SIN); (c) ground truth sinogram; (d) reconstruction result of state-of-art iterative SART method with TV normalization; (e) our method output; (f) ground truth image}
\label{fig:init_fig}
\end{figure}

Limited angle CT reconstruction is a very challenging ill-posed issue and of
great interest in several clinical applications, such as digital breast
tomosynthesis \cite{zhang2006comparative}, dental tomography
\cite{rantala2006wavelet}, short exposure time \cite{jin2010motion}, etc. In a
limited-angle CT scan, the projection data can be obtained in less than \(180
^\circ\) angular range, and the data insufficiency degrades reconstruction
quality with streaking artifacts(fig \ref{fig:init_fig} (a)(b)).

While the uniqueness of solutions for non-truncated limited-angle CT has been
proved by analytic continuation theory \cite{ye2008exact}, the academia has
not yet found a closed-form solution for the problem. A common approach to the
problem is an iterative reconstruction algorithm with regularization term based
on compressed sensing theory \cite{candes2006robust}. The academia has reported
some iterative algorithms, e.g. classic ART \cite{gordon1970algebraic}, POCS
\cite{candes2005signal}, and some regularization terms, e.g. TV
\cite{0031-9155-54-9-014}, ATV \cite{chen2013limited}. These algorithms can
suppress artifacts and improve image quality but require high computational cost
for iterative projections and back-projections.

Another attempt to tackle insufficient data reconstruction problem is sinogram
inpainting, which has been explored for recovering incomplete information in
projection, e.g. metal artifact reduction (MAR)
\cite{mehranian2011sparsity,chen2012ct,peng2017gaussian,kim2010effective} ,
sparse-view reconstruction based on dictionary learning \cite{li2014dictionary} ,
and suppresses ring artifacts caused by detector gaps \cite{li2012strategy}.
However, limited-angle sinogram inpainting remains challenging for large
missing regions, which is known as the "semantic hole-filling" issue. An exemplary
algorithm GP-EL \cite{gao2007fast}, which tries to recover the sinogram signal
based on a GP extrapolation algorithm, still failed to produce satisfying results.
Therefore, the iterative reconstruction remains the state-of-art method.

In recent years, deep learning(DL) \cite{lecun2015deep} has attracted a lot of
attention because of convolutional neural networks'(CNN) outstanding performance
in image classification \cite{he2016deep,szegedy2017inception}, object detection
\cite{redmon2016you,girshick2015fast,ren2015faster}, and super-resolution
\cite{kim2016accurate,ledig2016photo}. Meanwhile, Deepak Pathak proposed Context Encoders
\cite{pathak2016context}, the first network able to give reasonable results for
"semantic hole-filling" issue, which is a promising approach to solve the limited-angle
sinogram inpainting task.

Data-driven learning method has been explored in CT field, e.g. a redundant
 learned dictionary as a CS sparsity normalization term in sinogram inpainting for
 MAR \cite{li2014dictionary} and low-dose CT reconstruction \cite{xu2012low}. CNNs
 have been showing their huge potential ability in CT reconstruction. Zhang H.M
 \cite{zhang2016image} adopted a 3-layer deep neural network on post-processing
 limited-angle CT FBP reconstruction, with the benefits of artifacts reduction
 and image details recovery. Kyong H.J. et. \cite{jin2017deep} proved that
 iterative methods can be viewed as a CNN and proposed FBPConvNet,
 FBP reconstruction combined with a U-Net \cite{ronneberger2015u} CNN, for
 sparse-view CT reconstruction. Ge W.'s pilot experiment
 \cite{wang2016perspective} shows the potential of 3-layer deep network to improve
 the image and sinogram quality, e.g. inpaint in the sinogram for eliminating
 metal artifacts. Hu C. et. \cite{chen2017low} proposed a "Residual
 Encoder-Decoder Convolutional Neural Network" (RED-CNN) to eliminate image
 noise in low-dose CT.

Despite these works, two practical and theoretical questions remain regarding
applying deep learning method in limited-angle CT reconstruction. First one, the
network mentioned above focused on either image domain
\cite{jin2017deep,zhang2016image,chen2017low} or sinogram domain \cite{wang2016perspective},
without cross-domain information transferring during training. So an image domain
network may generate results not satisfying data fidelity, and a sinogram domain
network inpainting result may have small inconsistency can cause significant
artifact in reconstruction. To tackle this problem, cross-domain error
backward-propagation is in need, which requires new-designed differentiable
reconstruction and projection method embedded in the network. The second one is that
networks mentioned above were trained supervised, i.e. with labeled/paired
training data. However, labeled data are limited in amount or expensive
\cite{wang2016perspective} in the clinical world due to some reasons, like dose or
privacy.

In this paper, we start our discussion proving sinogram inpainting function is
continuous based on analytic continuation theory, so it can be fitted by the neural
networks in data-driven method guaranteed by universal approximation theorem
\cite{csaji2001approximation}. We propose Sinogram Inpainting Network(SIN), a
convolutional neural network trained to generate missing sinogram data conditioned
on projections from the scan, as the fitting function. To implement the cross-domain
error backward-propagation we propose Radon and Inverse Radon Transformer
network. This differentiable module can be inserted into existing convolutional
network architectures, which enables to perform CT projection and/or reconstruction
within the network. And we adopt an image processing network in our networks to
eliminate the artifacts caused by small inconsistency in the inpainted sinogram.
The three parts form an end-to-end network from sinogram domain to image domain, with
benefits of taking both of image error and sinogram error into account in the sync
process in supervised training, i.e with limited-angle/full-view sinogram
pairs.

To tackle this training data bottleneck, we develop an unsupervised train method
with only limited-angle projection on our proposed network. Inspired by the
observation: reconstruction and projection can form a closed loop, we can get a
fake projection from the reconstructed image, and the disparity between the
fake projection and real projection gives feedback signals to train proposed
network unsupervised.

We demonstrate the performance of our proposed network on lung CT datasets from
Data Science Bowl 2017, and the ability to unsupervised transfer learning on
Zubal's phantom \cite{zubal1994computerized}. The proposed method highly improves
limited-angle CT reconstruction quality and performs better than state-of-art
method iterative SART with ATV normalization in PSNR and SSIM quantitative
metrics. Visual improvements in our results are easily noticeable.

\section{Method}
\label{sec:orga4e1f3d}
In this chapter, we will cover why sinogram inpainting task can be implemented
by neural network, what architecture of neural network is adapted to the task
and how to train the neural network. We start by proving sinogram inpainting
task is a continuous function of original projection. Going on, we demonstrate
the three parts and their contributions in the proposed network:sinogram
inpainting network, radon/inverse radon transfer module and image processing
network. Finally, we defines the training objective and loss function for
supervised and unsupervised training.

\subsection{Theory}
\label{sec:orgb6c9a47}
\label{theory}

\begin{figure}[htb]
\centering
\subfloat[]{\includegraphics[width=2.6in]{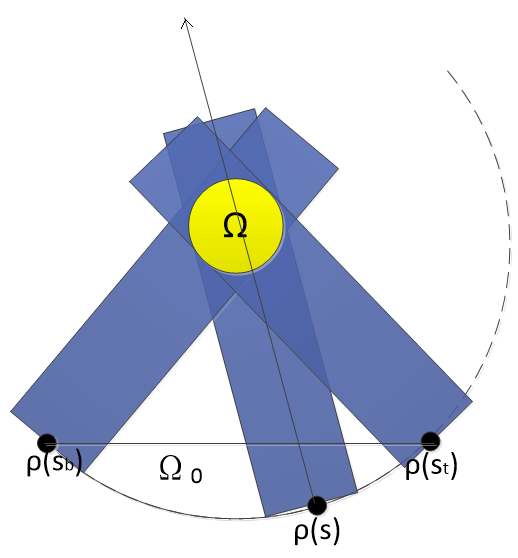}}\\
\subfloat[]{\includegraphics[width=3.3in]{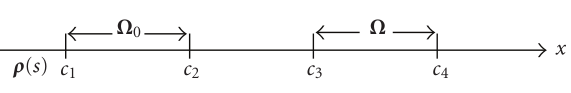}}
\caption{a) Illustration of parallel-beam limited angle scanning geometry, where solid line is the trajectory of source $\rho$ and dashed line represents the missing data; (b) 1D coordinate along the x-ray source path indicated in (a).}
\label{fig:theory}
\end{figure}

For the continuous nontruncated limited-angle problem, we use parallel-beam case
as a representative, illustrated in Figure \ref{fig:theory}, and the following
proof applies equally to fan-beam case. Object function \(f(\mathbf{x})\) is
constrained in a compact support \(\Omega \subset \mathbb{R}^2\), and \(\Gamma\) is
a general smooth source trajectory.

\begin{equation}
\Gamma = \{\rho(s) | s \in \mathbb{R} \}
\end{equation}

\begin{figure*}[tb]
\centering
\includegraphics[width=7in]{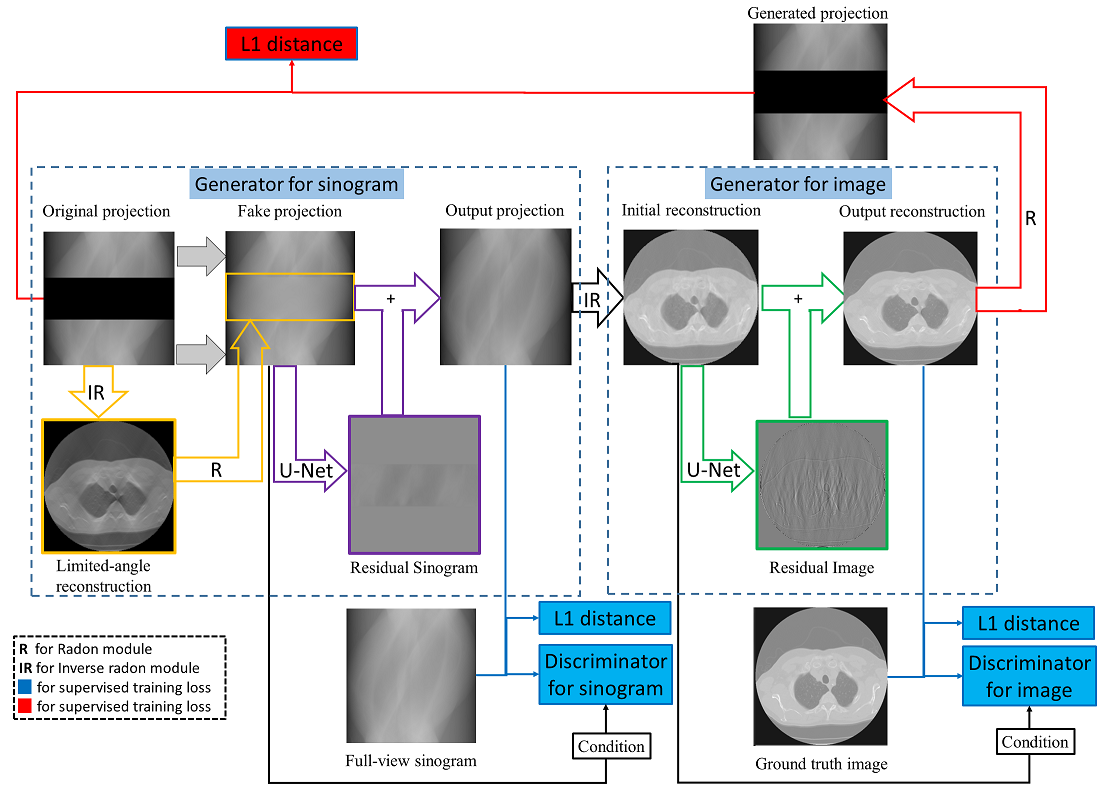}
\caption{Architecture of the proposed network. The $G_s$ is residual. The blue blocks denote the loss for supervised training and the red blocks denote the loss for unsupervised training. }
\label{fig:network_design}
\end{figure*}

Two points \(\rho(s_b)\) and \(\rho(s_t)\) are start and terminal points of the We
begin our discussion by proving describing our on continuous case. limited
scanning, \(\rho(s)\) is any point satisfying \(s_b<s<s_t\) , and the dashed line
represents missing data with point \(\rho(\tilde{s})\) on it. Assuming that
connecting \(\rho(s_b)\) and \(\rho(s_t)\) forms a measurable region \(\Omega_0\), we
can set 1D coordinate along any X-ray \((s,\theta)\) intersects \(\Omega\) from
\(\rho_s\), and we denote points where the X-ray intersects with \(\Omega_0\) and
\(\Omega\) as \(c_1, c_2, c_3, c_4\) , with \(c_1< c_2< c_3< c_4\)

Based on the inverse Hilbert Transform \cite{noo2004two}, we have the
reconstructiont formulation for \(f(x)\) from the research of Yangbo Y.
\cite{ye2008exact} , where \(g(x)\) is the 1D Hilbert transform of \(f(x)\)

\begin{equation}
\label{eq:1}
\begin{split}
&\sqrt{(c_4 -x)(x-c_1)}f(x)\\
&= \int_{c_1}^{c_2} \frac{\sqrt{(c_4 -\tilde{x})(\tilde{x}-c_1)}g(\tilde{x})}{\pi(\tilde{x} - x)} d\tilde{x} + \frac{1}{\pi}\int_{c_3}^{c_4}f(\tilde{x}) d\tilde{x}  \\
&+ \int_{c_2}^{c_4} \frac{\sqrt{(c_4 -\tilde{x})(\tilde{x}-c_1)}g(\tilde{x})}{\pi(\tilde{x} - x)} d\tilde{x} 
\end{split}
\end{equation}

Pack et al. \cite{pack2005cone} presents a local operation for converting
projections into values of \(g(\tilde{x})\). We use \(\mathcal{L}\) to represent the
coverting operator for clarity, and \(P\) to be the scanning projection data.

\begin{equation}
g(\tilde{x}) = \mathcal{L}(P,\tilde{x})  ~ , ~~ x \in (c_1,c_2)
\end{equation}

We denote the last part of equation \ref{eq:1} as \(G_{c_2 c_4}\). For \(f(x)|_{x \in (c_1,c_2)} = 0\)

\begin{equation}
\begin{split}
& G_{c_2 c_4}(x)|_{P,x \in (c_1,c_2)}  \\
&=   \int_{c_2}^{c_4} \frac{\sqrt{(c_4 -\tilde{x})(\tilde{x}-c_1)}g(\tilde{x})}{\pi(\tilde{x} - x)} d\tilde{x}  \\
&=  - \int_{c_1}^{c_2} \frac{\sqrt{(c_4 -\tilde{x})(\tilde{x}-c_1)}g(\tilde{x})}{\pi(\tilde{x} - x)} d\tilde{x}  - \frac{1}{\pi}\int_{c_3}^{c_4}f(\tilde{x}) d\tilde{x}  \\
&=  - \int_{c_1}^{c_2} \frac{\sqrt{(c_4 -\tilde{x})(\tilde{x}-c_1)}\mathcal{L}(P,\tilde{x})}{\pi(\tilde{x} - x)} d\tilde{x} - \frac{1}{\pi}P(s,\theta)
\end{split}
\end{equation}

As \cite{ye2008exact} pointed out, \(G_{c_2 c_4}(x)|_{P}\) is an analytic function
with a cut on \([c_2, c_4]\), so it can be extended to \([c_2, c_4]\) as a
continuous function, denoted as \(\mathcal{E}G_{c_2 c_4}|_{P}\). Then we can get f(x) following equation \ref{eq:1} 

\begin{equation}
\begin{split}
f(x)|_{P,x \in (c_3,c_4)} =\frac{\frac{P(s,\theta)}{\pi} + \mathcal{E}G_{c_2 c_4}(x)|_{P}
+L(x)|_P}{\sqrt{(c_4-x)(x-c_1)}} \\
L(x)|_P = \int_{c_1}^{c_2}\frac{\sqrt{(c_4 -\tilde{x})(\tilde{x}-c_1)}\mathcal{L}(P,\tilde{x})}{\pi(\tilde{x} - x)} d\tilde{x}
\end{split}
\end{equation}

We can use Radon Transform, denoted as \(\mathcal{R}\), to get the missing projection data \(\tilde{P}\).

\begin{equation}
\begin{split}
\tilde{P} &= \mathcal{R} f(x)|_P \\
&= \int_{\tilde{s},\tilde{\theta}}\frac{\frac{P(s,\theta)}{\pi} + \mathcal{E}G_{c_2 c_4}(x)|_{P} +L(x)|_P}{\sqrt{(c_4-x)(x-c_1)}} \mathrm{d} x \\
&= F(P)
\end{split}
\end{equation}

So we have proven the sinogram inpainting target \(\tilde{P}\) is a function of
scanning data \(F(P)\).Because \(\mathcal{L}(P,\tilde{x})\), \(G_{c_2 c_4}(P,x)\) and
\(\mathcal{E}\) are all continuous on \(P\), so the \(F(P)\) is continuous on \(P\).

Based on universal approximation theorem \cite{csaji2001approximation} , a
feed-forward network with a single hidden layer containing a finite number of
neurons can be used to approximate continuous functions on compact subsets of
\(\mathbb{R}^n\). As \(\mathbb{C}\) can be viewed as \(\mathbb{R}^2\), the conclusion
also applies to our function. So we propose a deep network trained to generate
missing sinogram data conditioned on projections from scan.

\subsection{Network Design}
\label{sec:org229ec89}
\begin{figure}[tb]
\centering
\includegraphics[width=3.3in]{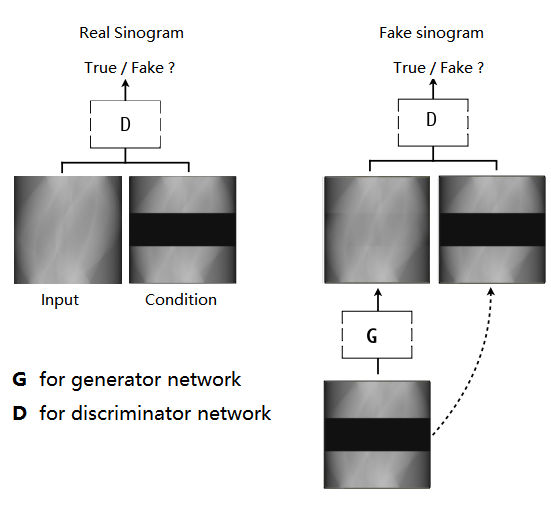}
\caption{\label{fig:D}
Sinogram inpainting network is a conditional GAN. The discriminator D learns to identify real and fake sinogram. The generator learns inpaint the sinogram to fool the discriminator. The limited angle projection is used twice as condition.}
\end{figure}

\begin{figure}[tb]
\centering
\subfloat[]{\includegraphics[width=3.3in]{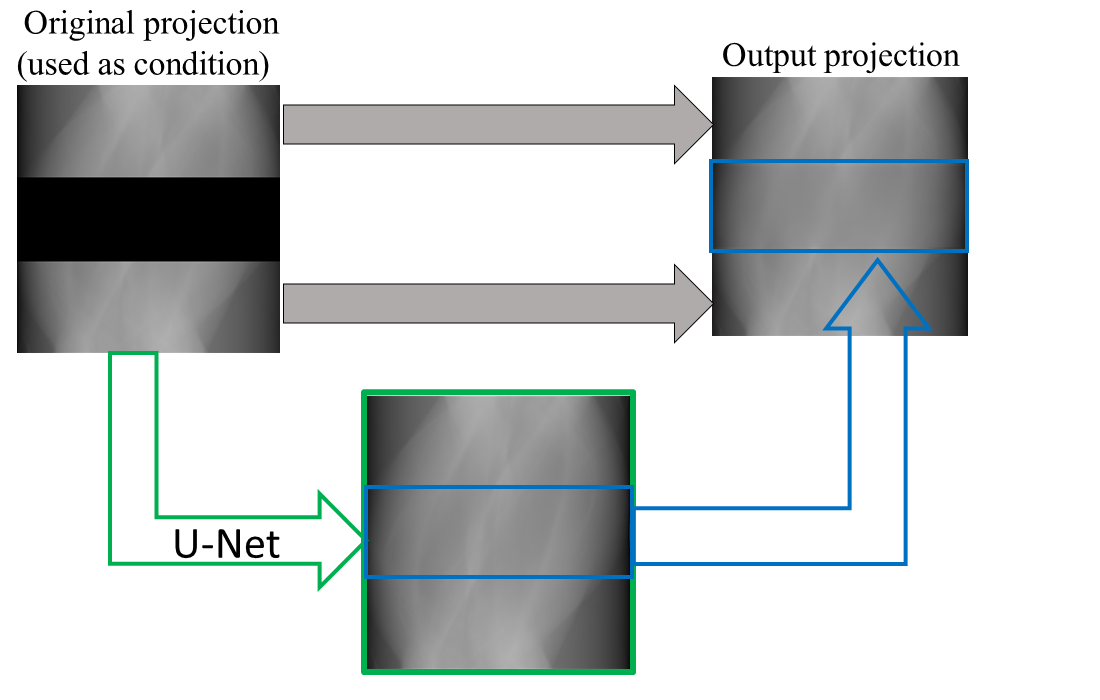}
}\\
\subfloat[]{\includegraphics[width=3.3in]{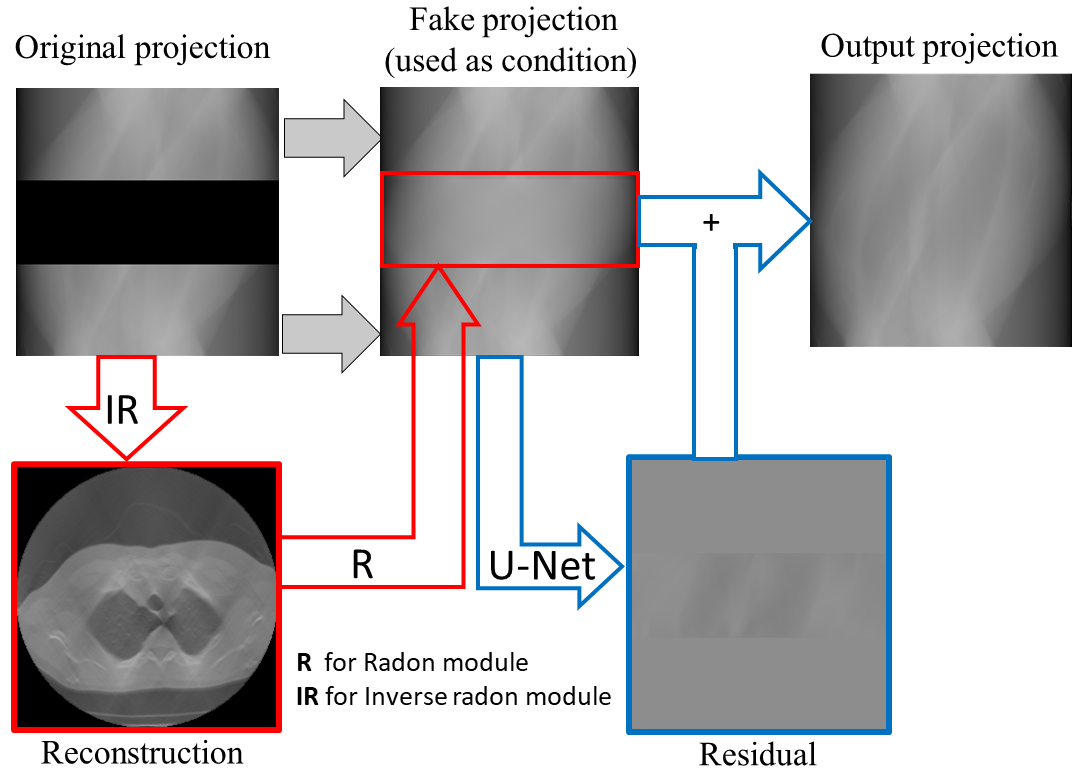}
}
\caption{two different architecture of the generator network for sinogram inpainting, (a) is called non-residual $G_s$ and (b) is the proposed one and called residual $G_s$.}
\label{fig:SIN}
\end{figure}

Based on the proof in \ref{theory}, it's possible to train an end-to-end neural
network from sinogram to reconstruction image, but requires the network to
encode the conversion from polar coordinate to the rectangular coordinates by
learning \cite{jin2017deep}. So we seperate the process into three parts: the
sinogram inpainting network (SIN, shown in fig \ref{fig:SIN}), radon/inverse
radon transformer module (shown in fig \ref{fig:radon}), and image processing
network. This configuration keeps SIN network only processing sinogram
information and so does image processing network in image domain, which
simplifies the network architecture and reduce learning complexity.

 The network architecture is shown in fig \ref{fig:network_design} and the
effect of each part is demonstrated in fig \ref{fig:Gi_results}. First we
introduce the network parts and its effect in supervised training. In the last
part, we will introduce the unsupervised training method.

\subsubsection{Sinogram Inpainting Network}
\label{sec:orga177668}

In this paper, we base our sinogram inpainting network on pix2pix
 network \cite{DBLP:journals/corr/IsolaZZE16}, concluding two parts: the generator
 \(G_s\) and the conditional discriminator \(D_s\), where \(G_s\) takes limited-angle
 sinogram as input and generates fake images that fool \(D_s\), and \(D_s\) tries to
 discriminate real and fake images based on the condition image. \(D_s\) is only
 used in supervised training.

We use pix2pix network to replace Context Encodes base network for two reasons:
i) pix2pix adds skip connection in the generator like "U-Net"
\cite{ronneberger2015u}, so the low-level feature can be shuttled directly, which
is a good analog for multiresolution single reconstruction. ii) pix2pix uses
conditional adversarial networks instead of traditional adversarial network( shown in \ref{fig:D}), so
it can learn a conditional generative model with a conditional loss function
based on the input condition when we condition on the input sinogram in our
case.

As a refinement of pix2pix, we use the original data to replace corresponding
part in fake images, which forces the \(G_s\) focusing on learning the missing
part. And we compairs two different types of generator(as shown in
\ref{fig:SIN}), one tries to generate the missing data from the original data
directly, which is called non-residual \(G_s\). The other one reconstructs image
with limited-angle arifacts and generates the fake projection data for missing
angle firstly, and try to learn the difference between the fake projection and
ground truth. We can see the residual \(G_s\) performs better in experiment
(shown in fig \ref{fig:SIN_results}) , for reconstruction and projection
process encapsulates the physical model and preserves some information.

\subsubsection{Radon and inverse radon module}
\label{sec:orgb3721d1}
\label{radon}

\begin{figure}[tb]
\centering
\subfloat[]{\includegraphics[width=3.3in]{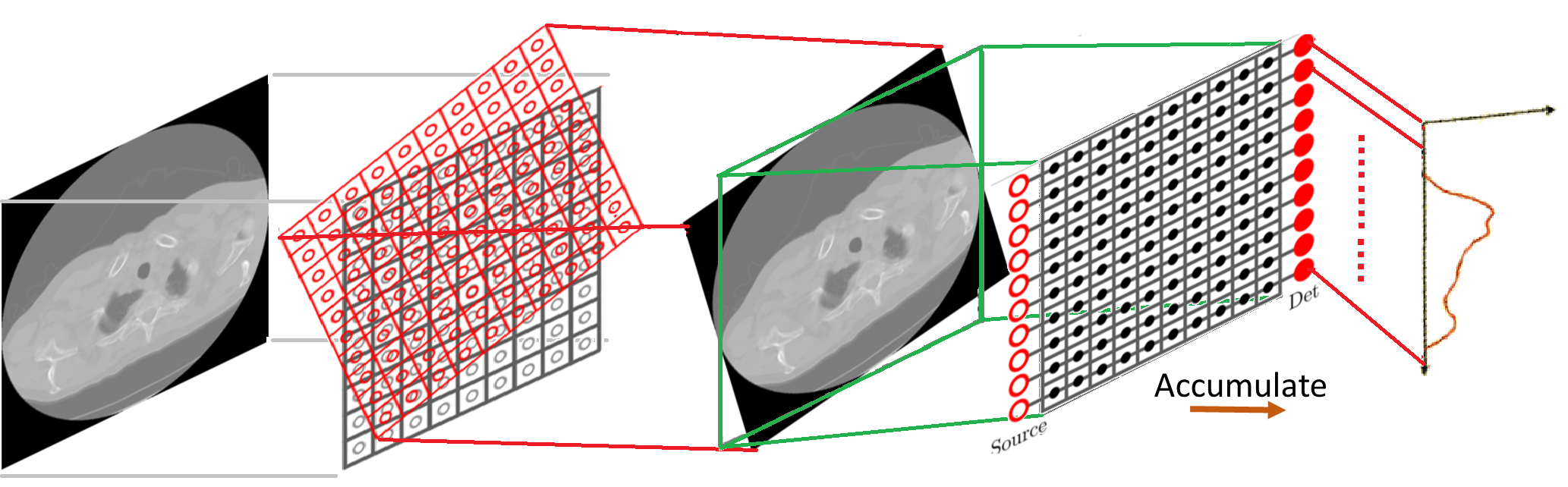}}
\\
\subfloat[]{\includegraphics[width=3.3in]{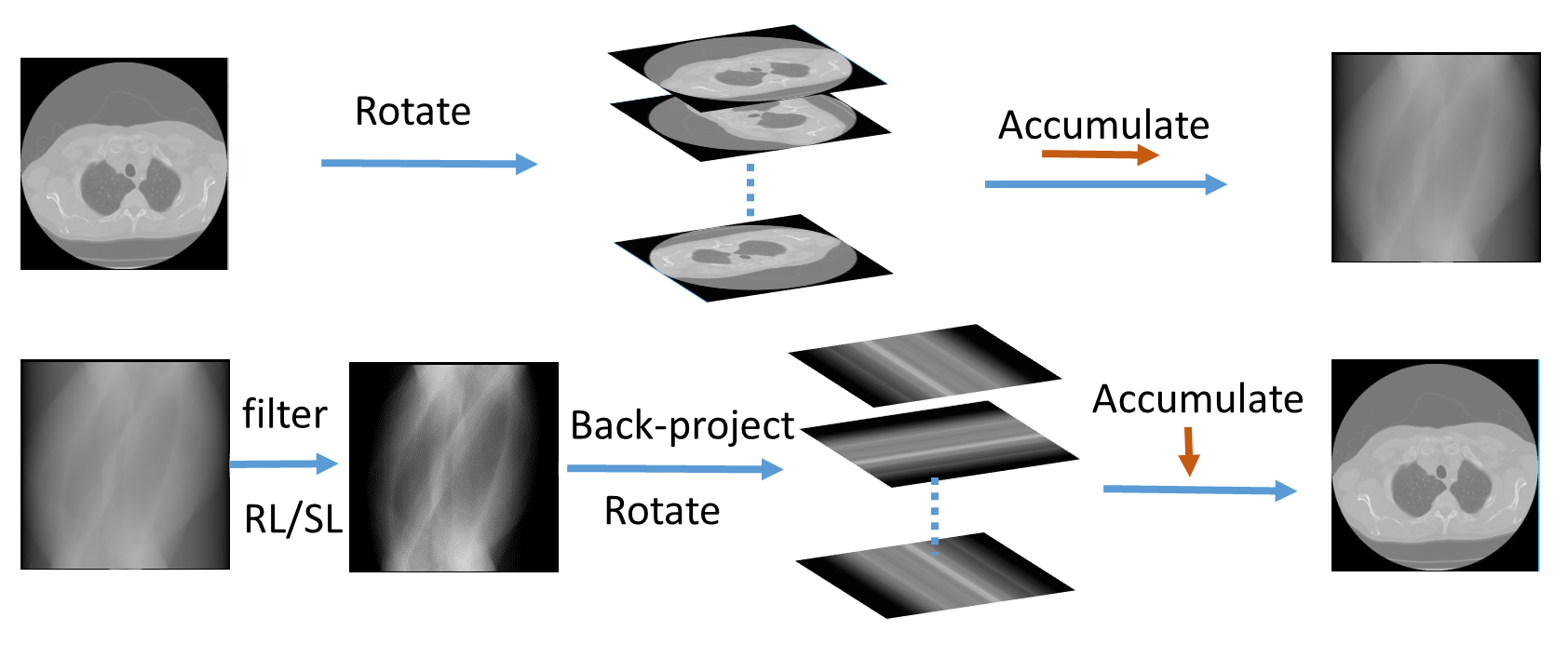}
}\\
\subfloat[]{\includegraphics[width=2.8in]{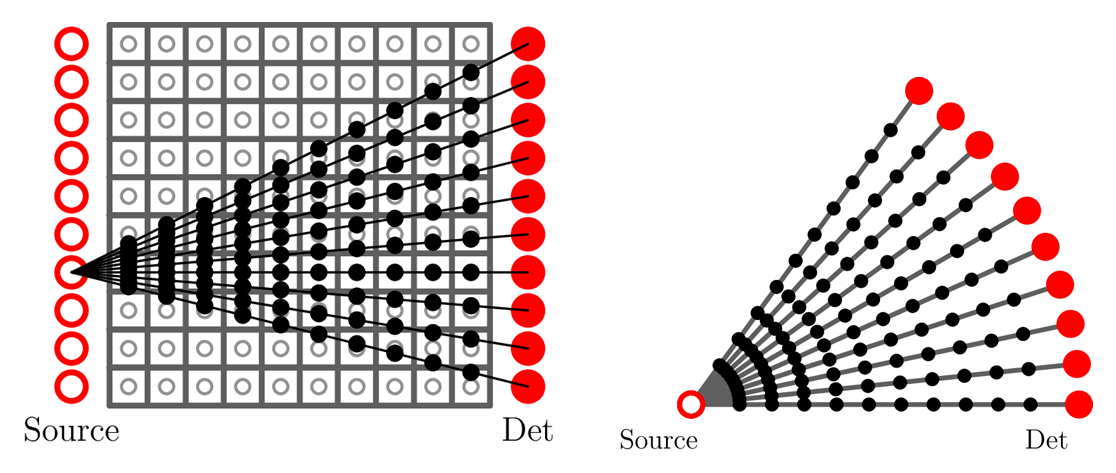}
}
\caption{Illustration of Radon Transformer and Inverse Radon Transformer network(a) The interpolation and accumulation process; (b) The Radon transformer can be viewed as rotate interpolation and accumulation. The inverse Radon transformer can be viewed as: 1D filtering(SL, RL ,etc.), interpolation, and accumulation; (c) This module can be easily extended to other geometry, e.g. equal-distance and equal-angle fan-beam.}
\label{fig:radon}
\end{figure}

As we all know, minor inconsistencies in sinogram may lead to serious artifacts
in the reconstructed image. It's benefical to take both of image error and
sinogram error in account in training, which requires a reconstruction module
with the ability for error backward-propagation from image domain to sinogram
domain. So iterative methods are not suitable.

And it's challenging to deploy FBP reconstruction method in neural network,
because traditional implement of FBP requires the mulitplication of huge system
matrix. It will take too much space if stored in dense array, (for a \(256*256\)
reconstruction image detected by \(256\) detector pixels in \(360\) views, the array
size is 24.16GB in float32 type , the sparsity is approximate \(0.0078\)). On the
other hand, compution on sparse array is time-consuming for forward and backward
propagation.

We propose a differentiable Radon transformer network \(\mathcal{R}\) and inverse
Radon transformer network \(\mathcal{\tilde{R}}\) with the ablity or error
backward-propagation, to replace system matrix mulitplication(shown in fig
\ref{fig:radon}). The Radon transformer,i.e. parallel-beam projection, can be
seperated by two parts: interpolation (inspired by spatial transformer network
\cite{jaderberg2015spatial}), and accumulation. The inverse Radon transformer, i.e.
parallel-beam reconstruction, concluded: 1D filtering(SL, RL ,etc.),
interpolation, and accumulation. The system matrix coefficients are replaced by
densely stored interpolation coefficients. This module can be easily extended to
other geometry, e.g. fan-beam, cone-beam.

This differentiable module can be inserted into existing convolutional
architectures, which enables to perform CT projection and/or reconstruction
within the network and the optimisation process.

\subsubsection{Image processing network}
\label{sec:org76d6024}

As mentioned before, some reconstruction artifacts are corresponding to minor
inconsistencies in sinogram. The idea is natural to eliminate these arifacts in
image domain more easily than in sinogram domain, so it should be benefical to
have an deep network for image processing. We still base image processing
network on pix2pix, where the generator is denoted as \(G_i\), the discriminator
is \(D_i\), and FBP reconstruction image from original limited angle sinogram as
condition. \(D_i\) is only used in supervised training.

As a modification, we add a skip connection between \(G_i\) input and output,
like \cite{jin2017deep,kim2016accurate}. So \(G_i\) only learns the residual part
\(\mathbf{r} = \mathbf{y} - \mathbf{x}\), which release the network from
preserving all information. At the same time, the skip connection makes the
\(G_s\) easier to optimize by providing a short-cut way for error
backward-propagation and tackles the vanishing/exploding gradients problem ,
like in Resnet \cite{he2016deep}.

The generator part of this network is very similar to \cite{jin2017deep}, but we
add discriminator and adversarial loss, which is benefical for restore
high-frequency structure \cite{pathak2016context,DBLP:journals/corr/IsolaZZE16},
\subsubsection{Unsupervised transfer learning}
\label{sec:orgdfffcbc}

In some practical scene, paired training data, i.e. limited-angle projection and
full-view projection/ground truth image, is limited in amount or very costly.
\(D_s,D_i\) can't work without paired data, which limits the applicable scene for
the supervised training method. To tackle training data bottleneck, we develop a
unsupervised transfer training method with only limited angle projection.

Inspired by dual-learning \cite{he2016dual}, CT imaging can be view as a dual
task: reconstruction and projection can form a closed loop. With only limited
angle projection sinogram, we can reconstuct an image using \(G_s\), inverse radon
transformer and \(G_i\). So we can project reconstructed image by radon transformer
network, and the disparity between the generated projection and original
projection data gives feedback signals to train proposed network unsupervisedly.
The objective shown as red block in fig \ref{fig:network_design}.

\subsection{Objective}
\label{sec:org1ee181a}
\label{objective}

In supervised training case, our Loss function objective is the weighted sum of
cGAN objective and a traditional L1 distance, suggested by
\cite{pathak2016context,DBLP:journals/corr/IsolaZZE16} . The objective of a
conditional GAN can be expressed as \cite{pathak2016context},

\begin{equation}
\begin{split}
\mathcal{L}_{c,G_s}(G_s,D_s) = &\mathbb{E}_{x,y \sim p_s(x,y)}[logD_s(x,y)] \\
                            +\mathbb{E}_{x,y \sim p_s(x,y)}[ & log(1-D_s(x,G_s(x))] \\
\mathcal{L}_{c,G_i}(G_i,D_i) = &\mathbb{E}_{x,y \sim p_s(x,y)}[logD_i(\mathcal{\tilde{R}}(G_s(x)),\mathcal{\tilde{R}}(y))] \\
                               + \mathbb{E}_{x,y \sim p_s(x,y)}&[log(1-D_i(\mathcal{\tilde{R}}(G_s(x)),G_i(\mathcal{\tilde{R}}(G_s(x))))] 
\end{split}
\end{equation}

where \(p_s\) denotes data distribution, x is limited-angle sinogram data and y is
full-view data. \(D_s,D_i\) tries to discriminates real and fake data by
maximizing the loss and \(G_s,G_i\) tries to fool the discriminator by minimizing
the loss, i.e. \(G_s^*,G_i^* = \arg \min _{G_i,G_s} \max_{D_i,D_s}
[\mathcal{L}_{c,G_s}(G_s,D_s) + \lambda_i \mathcal{L}_{c,G_i}(G_i,D_i)]\)

The \(L_1\) loss plays the role like data fidelity items in image reconstuction,
keeping the generated data close the ground truth.
\begin{eqnarray}
&\mathcal{L}_{L1,G_s}(G_s) & = \mathbb{E}_{x,y \sim p_s(x,y)}[||y-G_s(x)||_1]  \\
&\mathcal{L}_{L1,G_i}(G_i) & = \mathbb{E}_{x,y \sim p_s(x,y)}[||\mathcal{\tilde{R}}(y)-G_i(\mathcal{\tilde{R}}(x))||_1]
\end{eqnarray}

The final loss fuction and objective are
\begin{equation}
\begin{split}
\mathcal{L} =&\lambda_i( \mathcal{L}_{c,G_i}(G_i,D_i) + \lambda_L \mathcal{L}_{L1,G_i}(G_i))\\
&+ \mathcal{L}_{c,G_s}(G_s,D_s) + \lambda_L \mathcal{L}_{L1,G_s}(G_s)) \\
G_s^*,G_i^* =& \arg \min _{G_i} \max_{D_i} \min_{G_s} \max_{D_s} \mathcal{L}
\end{split}
\end{equation}

For unsupervised transfer training case,
\begin{equation}
\begin{split}
\mathcal{L} = &\mathbb{E}_{x \sim p_s(x)}||(1-M)\odot(x-\mathcal{R}G_i\mathcal{\tilde{R}}G_s x)||_1\\
G_s^*,G_i^* =& \arg \min _{G_i} \min_{G_s} \mathcal{L}
\end{split}
\end{equation}

where \(\odot\) is the element-wise product operation, \(p_s\) denotes data
distribution, \(x\) is limited-angle sinogram data and \(M\) is limited-angle mask.
The loss plays the role of data fidelity items and aims to eliminate data
inconsistencies.

\section{Results}
\label{sec:orgb9707c9}
In this chapter, we will introduce our datasets, experiment setup and results.
In this paper, our experiments focus on parallel CT, but the method is general
and can be easily modified to apply in fan-beam CT as metioned before. We
compair the proposed method to FBP alone and state-of-the-art iterative
reconstrucion method with TV normalization \cite{0031-9155-54-9-014}, denoted as
SART-TV.

\subsection{Datasets and Experiment Configuration}
\label{sec:orgc5087d7}
We demonstrate the performance of our proposed network on two datasets: lung CT
datasets(fig \ref{fig:supervised_results}) from Data Science Bowl 2017 for
supervised training and test, and

Zubal's phantom CT data (fig \ref{fig:supervised_results}) from neck to
mid-thigh, except lung position, for test the ability of unsupervised transfer
learning. The lung CT train set is with 6312 images (from 38 persons) and test
set is with 113 images(from 2 persons). The Zubal's unsupervised test set is 40
images of 1 person.

Our data preparation goes as follows: firstly normlize the input image value
range to be [-1, 1] and image size to be \(256\times256\), and take this as ground
truth image. The we generate the projection using \(\mathcal{R}\) for 256 views in
\(180\deg\) with 256 detectors with the same size of image pixel, taken as ground
truth projection. The we cut the middle \(60\deg\) projection out(86 views), taken
as the limited-angle projection input. 

We use the Pytorch toolbox (ver. 0.2.0) to implement the proposed network
including SIN. And we use a Tesla P100 graphic card (NVIDIA Corporation) for
train and test. The hyperparameters for training are as follows: we use the adam
optimizer of all of \(G_i,G_s,D_i,D_s\), learning rate decreasing linearly from
0.002 to 0, the momentum \(\beta_1\) equals 0.5, \(\beta_2\) 0.999; batchsize is 30.
The SART hyperparameters are tuned manually to ge the best performance in
average, because it's the common practice in general and the academia haven't
come to the common idea of data-driven parameters setting method. We finally set
the 60 iteration for each image, and in each epoch we take 20 steps of TV with
the factor \(\alpha = 0.06, \alpha_s = 0.997\).

\subsection{Comparsion With Different Architecture}
\label{sec:orgd904808}

We take two samples from lung CT test set to demonstrate the difference
corresponded to changes in network architecture.

\begin{figure}[tb]
\centering
\subfloat[]{\includegraphics[width=1.1in]{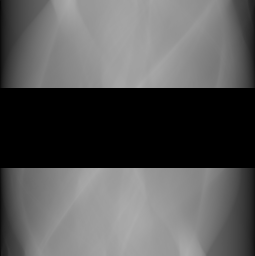}}
\hfill
\subfloat[]{\includegraphics[width=1.1in]{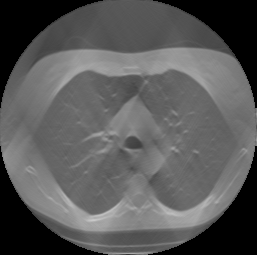}}
\hfill
\subfloat[]{\includegraphics[width=1.1in]{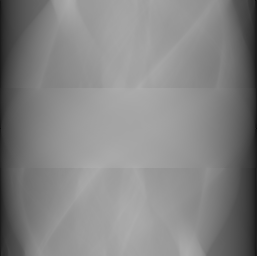}}
\\
\subfloat[]{\includegraphics[width=1.1in]{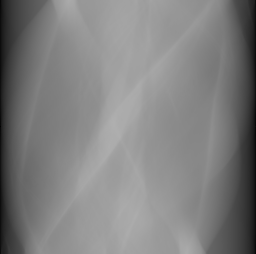}}
\hfill
\subfloat[]{\includegraphics[width=1.1in]{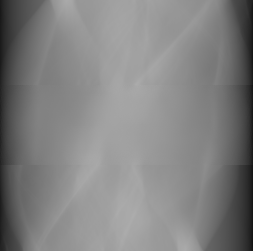}}
\hfill
\subfloat[]{\includegraphics[width=1.1in]{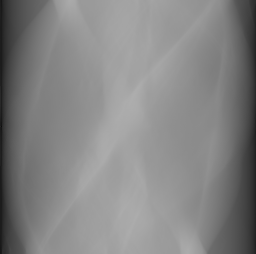}}
\\
\subfloat[]{\includegraphics[width=1.1in]{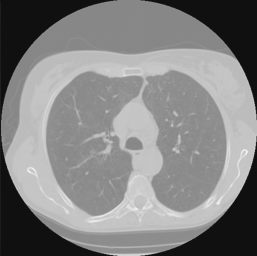}}
\hfill
\subfloat[]{\includegraphics[width=1.1in]{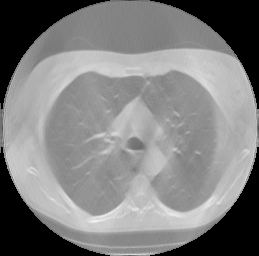}}
\hfill
\subfloat[]{\includegraphics[width=1.1in]{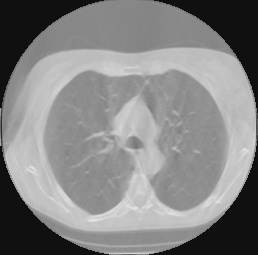}}
\\
\caption{Demonstrates inpainting results for different $G_s$ architecture a) the original limited-angle data; b) FBP reconstruction image for a); c) use projection from b) to replace the missing data in a);  d) the ground truth of sinogram;e) the output of non-residual $G_s$ with a) as input; f) the output of residual $G_s$ with c) as input;g) the ground truth of reconstruction image; h) FBP reconstruction for e); i) FBP reconstruction for f).}
\label{fig:SIN_results}
\end{figure}

Firstly, the inpainting results of non-residual/residual sinogram generator work
is shown in fig \ref{fig:SIN_results}. And the sinogram of non-residual \(G_s\)
has much more inconsistence than residual one as non-residual needs to keep more
information through network.

\begin{figure}[tb]
\centering
\subfloat[]{\includegraphics[width=1.1in]{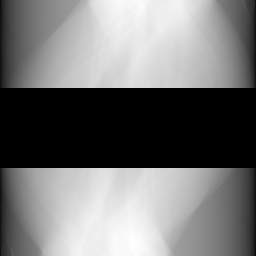}}
\hfill
\subfloat[]{\includegraphics[width=1.1in]{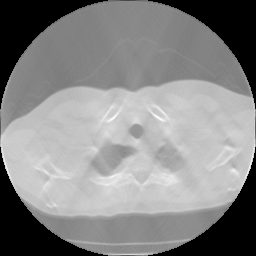}}
\hfill
\subfloat[]{\includegraphics[width=1.1in]{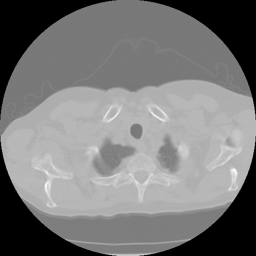}}
\\
\subfloat[]{\includegraphics[width=1.1in]{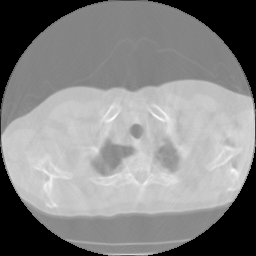}}
\hfill
\subfloat[]{\includegraphics[width=1.1in]{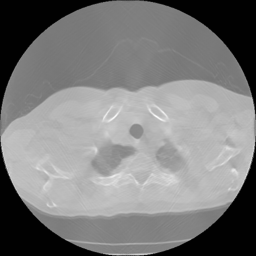}}
\hfill
\subfloat[]{\includegraphics[width=1.1in]{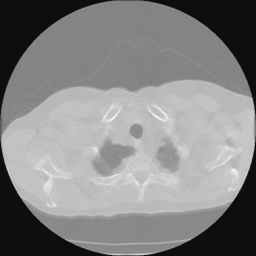}}
\caption{Demonstrates reconstruction results eliminating different part in our Network. a) the original limited-angle data; b) FBP reconstruction image for a); c) the ground truth of reconstruction image;  d)  Eliminate the $G_i$ part, i.e. reconstruction from $G_s$ output; e) Eliminate $G_s$, and train $G_i$ with b) as input, like \cite{jin2017deep} ; f) the output of residual $G_s$ with c) as input.}
\label{fig:Gi_results}
\end{figure}

Secondly, comparison of reconstruction results eliminating different part in our
Network is shown in fig \ref{fig:Gi_results}, which helps demonstrate the
individual contribution of the elements. For example, comparing fig
\ref{fig:Gi_results} (d) and (f) tells us the \(G_i\) can eliminate the artifacts
due to sinogram inconsistence so small to tackle, as we imagined. Result for
eliminating the \(G_s\) and taking the FBP reconstructed images as \(G_i\) input,
shown in fig \ref{fig:Gi_results} (e) , which is similar to the architecture in
\cite{jin2017deep} , suffers image quality degradation due to the blurring and
artifacts introduced by the filtering process in FBP.

\subsection{Quantitative Metrics and Experimental Results}
\label{sec:orgd34eedf}
We choose two quantitative metric to measure the quality of reconstruction: peak
signal to noise ratio (PSNR), and the mean structural similarity index (SSIM)
with the ground truth images. If \(x\) denotes the ground truth \(\hat{x}\) denotes
the algorithm output, PSNR and SSIM are given by

\begin{eqnarray}
&\hbox{PSNR}(\hat{x},x)=20\cdot \log \frac{\mathcal{R}_x}{\sqrt{\mathrm{E}(\hat{x}-x)^2}} \\
&\hbox{SSIM}(\hat{x},x)={\frac  {(2\mathrm{E} _{\hat{x}}\mathrm{E} _{x}+c_{1})(2\sigma _{{\hat{x}x}}+c_{2})}{(\mathrm{E} _{\hat{x}}^{2}+\mathrm{E} _{x}^{2}+c_{1})(\sigma _{\hat{x}}^{2}+\sigma _{x}^{2}+c_{2})}}
\end{eqnarray}
where \(\mathcal{R}_x\) denotes the range of x, i.e. \(x_{\max} -x_{\min}\),
\(\mathrm{E}\) denotes the average, \(\sigma^{2}\) denotes the variance. \(\sigma
_{xy}\) denotes the covariance of \(x\) and \(y\), \(c_1=(0.01 \mathcal{R}_x)^2\),
\(c_2=(0.03 \mathcal{R}_x)^2\) .It's easy to see PSNR estimates absolute errors
and SSIM is more sensitive for changes in structural information. Both higher
PSNR and SSIM values correspond to better reconstruction.

\subsubsection{Lung CT Test datasets}
\label{sec:orgdfba13c}

\begin{figure*}[htb]
\centering
\includegraphics[width=7in]{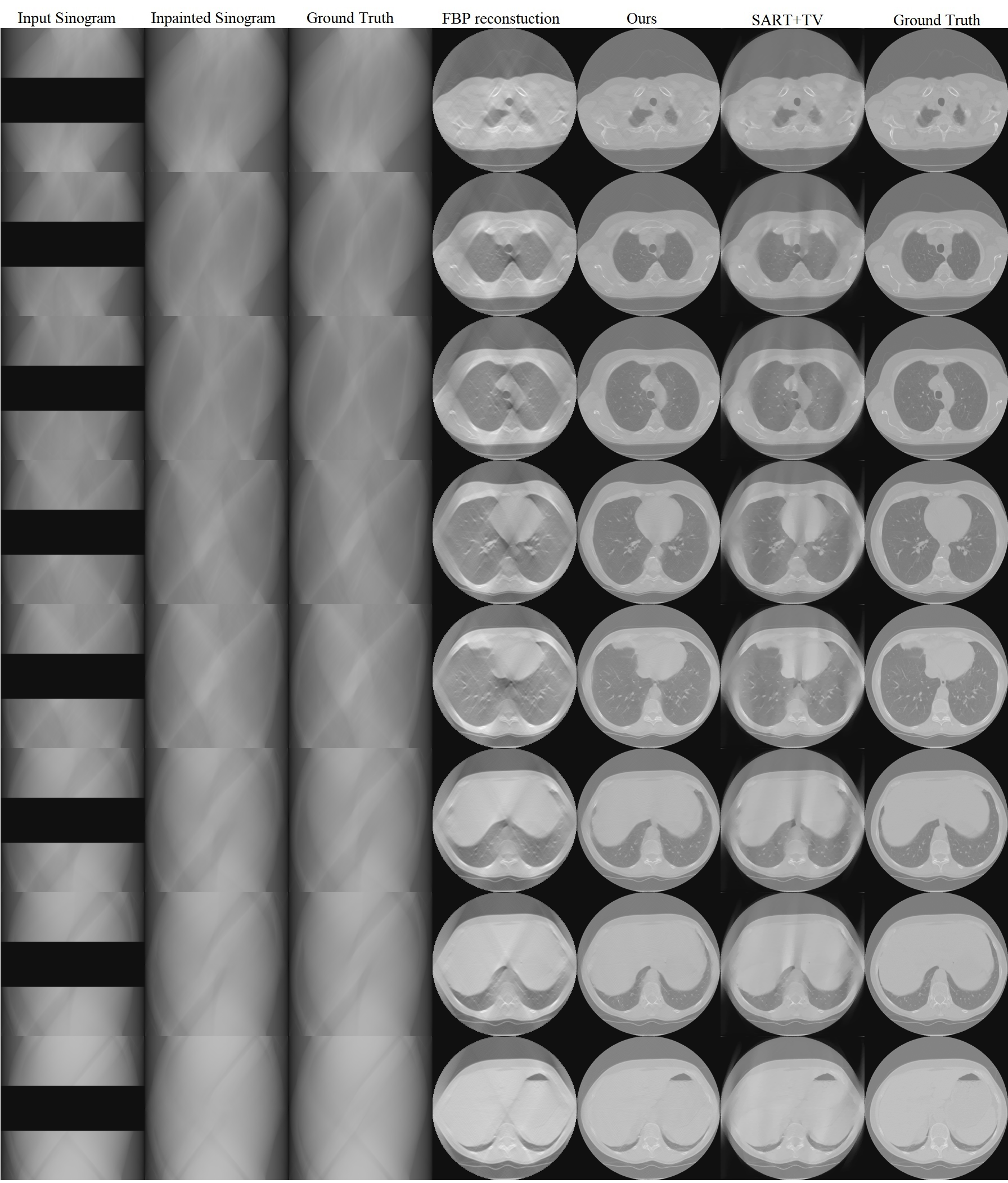}
\caption{The inpainted sinogram and reconstructed images for test datasets of lung CT from Data Science Bowl 2017. We compair three reconstrucion methods: FBP, our proposed supervised-trained network, and iterative SART reconstrucion with TV regularization. Both the proposed network and SART-TV method reduce the streaking artifacts. And our results have less cartoon-like artifacts than SART-TV. The display window is [-1, 1].}
\label{fig:supervised_results}
\end{figure*}

Fig \ref{fig:supervised_results} and Table \ref{tab:supervised_results}
demonstrate the experiment results for the lung CT data. Both the proposed
network and SART-TV method reduce the streaking artifacts caused by the
sinogram inconsistencies in angle, while SART-TV introduced piecewise constant
cartoon-like artifacts because of the TV normalization. The quantitative
studies indicate the superiority of the proposed method in terms of absolute
error and structural.

\begin{table}[htb]
\renewcommand{\arraystretch}{1.3}
\caption{Quantitative Comparsion for FBP, SART-TV and Proposed Method for Lung CT test dataset (113 images)}
\label{tab:supervised_results}
\centering
\begin{tabular}{ccccc}
\hline\hline
&\bfseries FBP & \bfseries SART-TV & \bfseries Proposed \\ 
\hline\hline
avg. \bfseries PSNR(dB) & 22.51&25.23 &\bfseries 33.86\\
\hline
avg. \bfseries SSIM & 0.9249 &0.9088 &\bfseries 0.9781\\
\hline\hline
\end{tabular}
\end{table} 

\subsubsection{Unsupervised Learning}
\label{sec:orgd331a2e}
\begin{figure*}[htb]
\centering
\includegraphics[width=7in]{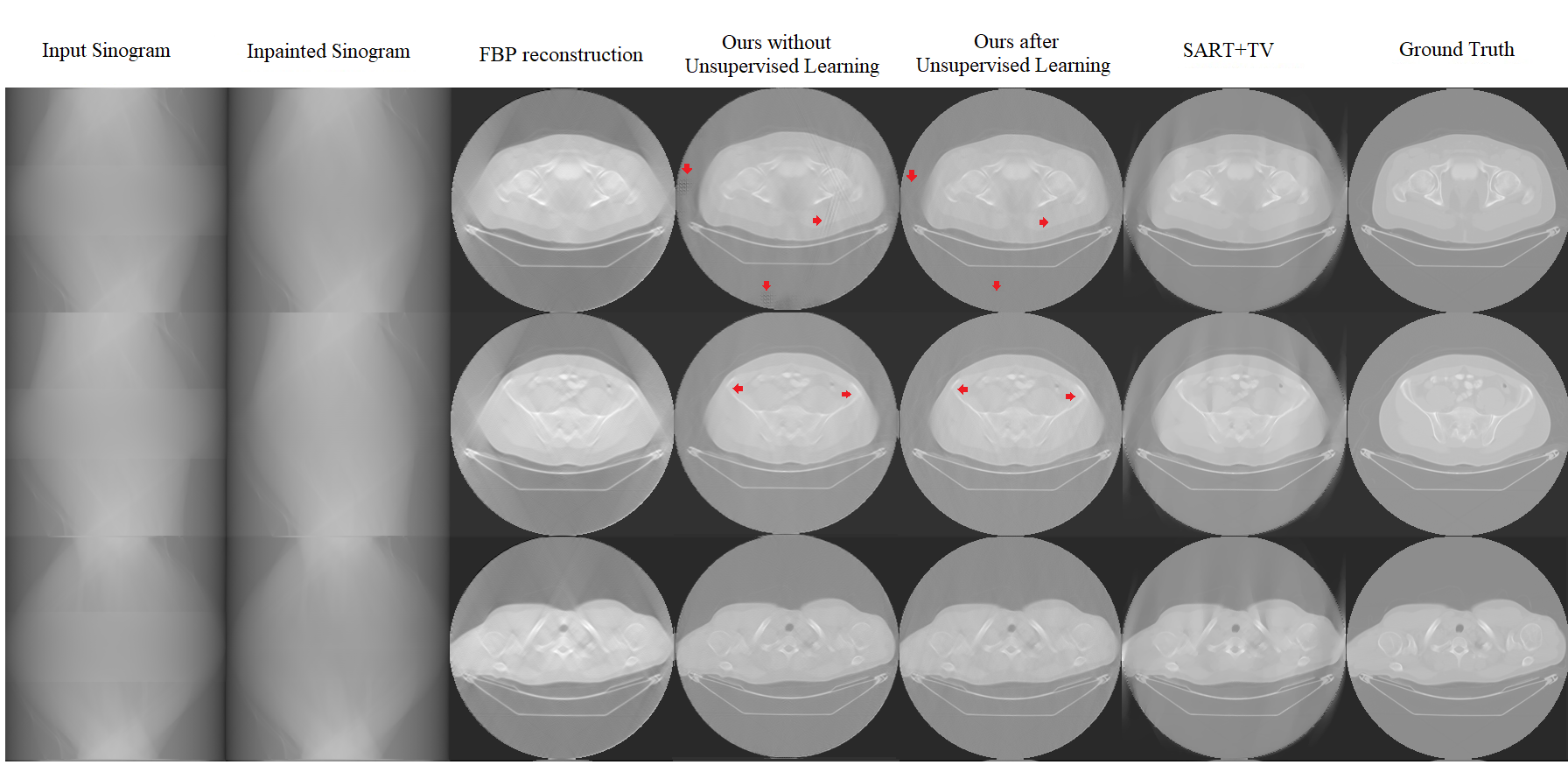}
\caption{The inpainted sinogram and reconstructed images for test datasets of Zubal's phantom. We compare three reconstruction methods: FBP, our proposed supervised-trained network before/after 30-epochs unsupervised learning, and iterative SART reconstruction with TV regularization. The results indicate the effects of unsupervised train, which consist of eliminating the artifacts introduced by overfitting (row1) and making the edges in image slightly sharper (row2). And the data fidelity implied in loss ensures the image quality not to degrade in training (row3). The display window is [-1, 1].}
\label{fig:unsupervised_results}
\end{figure*}

To validate and evaluate the generalization of the proposed method, we adopt the
network, trained on lung CT data, on the Zubal's phantom except lung position.
Fig \ref{fig:unsupervised_results} and Table \ref{tab:unsupervised_results}
demonstrate the experiment results.

We can see the process unsupervised training have significant effect in
 eliminating the artifacts introduced by over fitting
 (fig \ref{fig:unsupervised_results} row1) and make the edges in image slightly
 sharper (fig \ref{fig:unsupervised_results} row2). And the data fidelity implied in
 loss ensures the image quality not to degrade in training
 (fig \ref{fig:unsupervised_results} row3). Like the results
 for lung CT data, our after trained network's results have less streaking
 artifacts than FBP results and cartoon-like artifacts than SART-TV results. The
 quantitative metrics gives the same conclusion. So the after train proposed
 network can be transfer trained flexibly in different scanning scenes.

\begin{table}[htb]
\renewcommand{\arraystretch}{1.3}
\caption{Quantitative Comparsion for FBP, SART-TV and Proposed Method for Zubal CT dataset (40 slices)}
\label{tab:unsupervised_results}
\centering
\begin{tabular}{ccccc}
\hline\hline
&\bfseries FBP & \bfseries SART-TV & \bfseries Proposed \\ 
\hline\hline
avg. \bfseries PSNR(dB) & 24.88 & 28.08 & \bfseries 36.33 \\
\hline
avg. \bfseries SSIM & 0.9598 & 0.9523 & \bfseries 0.9838 \\
\hline\hline
\end{tabular}
\end{table}

\section{Conclusions}
\label{sec:orgac6f5c6}

In this paper we designed a network architecture with a deep convolutional
network SIN for sinogram inpainting task, new-designed differentiable
Radon/Inverse Radon Transformer modules for cross-domain error
backward-propagation and image processing network for eliminating artifacts.
Though we focused on limited angle task, the proposed sinogram inpainting method
can also be used in metal artifact reduction.

This approach can be trained both supervisedly and unsupervisedly to tackle the
labeled data bottleneck. In supervised training, our method can join both
sinogram and reconstruction error in the loss and optimize in sync with data
fidelity, thanks to the differentiable radon transformer. In unsupervised
training, we design a closed loop and use the inconsistence in closed loop as
loss function, so we can train on only limited-angle sinogram while satisfying
data-fidelity.

The proposed method performs better than state-of-art iterative reconstruction
SART-TV at different body positions, covering both supervisedly and
unsupervisedly learned ones. Furthermore, free from difficulty in selecting
superparameters and iterative projections/backprojections in SART, the
reconstruction time of our proposed network can be less than 0.3 second per
256*256 image in parallel batches.

\section{Appendix}
\label{sec:org0064993}

For someone that is interested in implementation in our proposed method, we
attach an architecture figure of modified pix2pix network for sinogram
inpainting network. And the pix2pix network in image processing network shares
the same architecture.

\begin{figure*}[htb]
\centering
\includegraphics[width=7in]{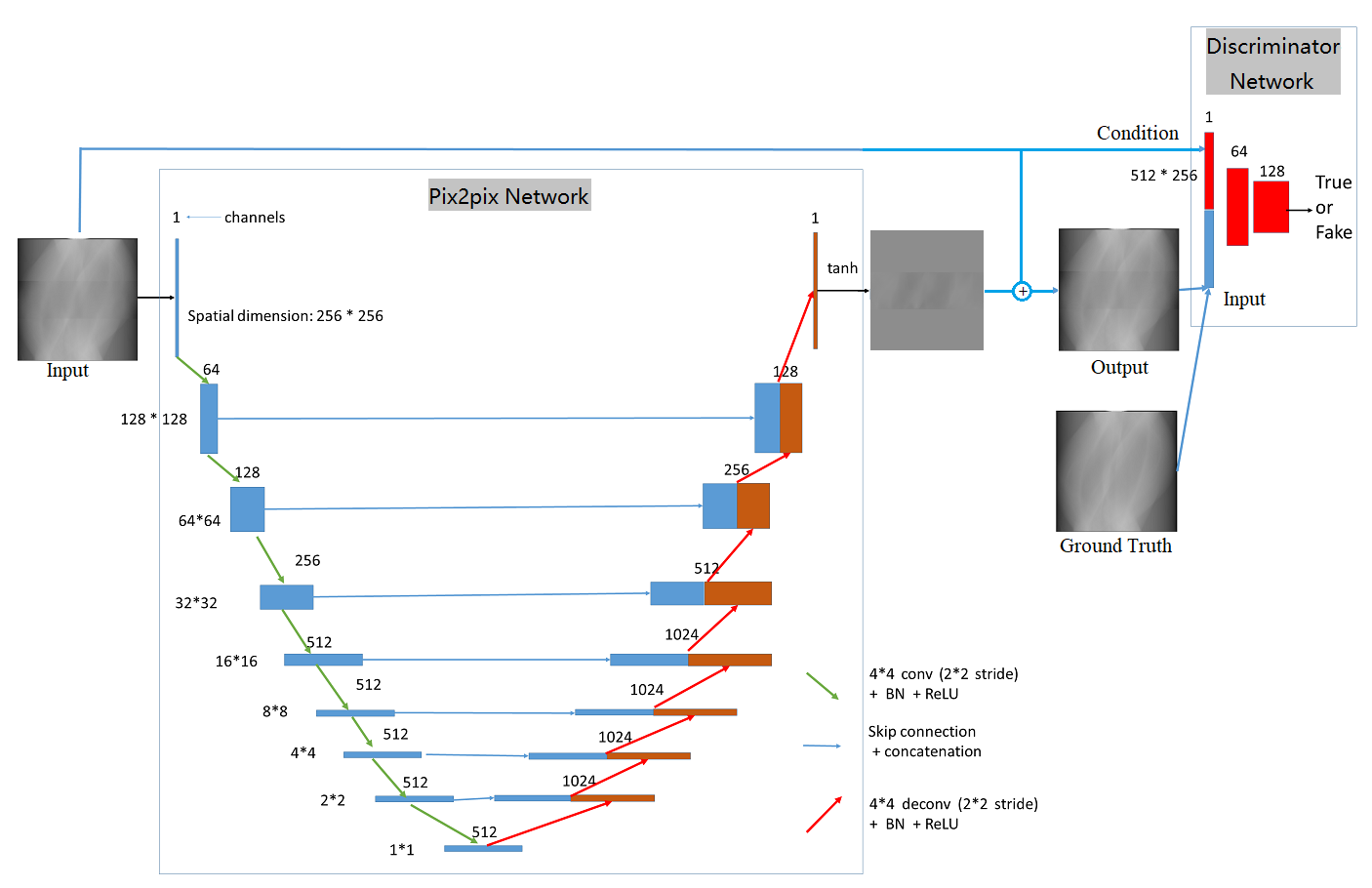}
\caption{Architecture of the proposed deep convolutional network based on pix2pix network except using the scanning data to replace corresponding part in fake images,
 which forces the $G_s$ focusing on learning the missing part.
}
\end{figure*}

\bibliographystyle{IEEEtran}
\bibliography{library}
\end{document}